%% file: CIAC-camera_ready.tex
\documentclass[runningheads]{llncs}
\usepackage{graphicx}
\usepackage[utf8]{inputenc}
\usepackage[T1]{fontenc}
\usepackage{paralist}
\newif\ifComments
\Commentsfalse
\usepackage{xspace}
\usepackage{color}
\usepackage[usenames,dvipsnames,svgnames,table]{xcolor}
\usepackage{tikz}
\usetikzlibrary{decorations.markings}

\usepackage{mathrsfs}
\usepackage{amsmath, amsfonts, amssymb, latexsym}
\usepackage{mathtools}
\usepackage{thm-restate}

\usepackage{url}
\usepackage{multirow}

\usepackage{float}
\newfloat{Algorithm}{t}{lop}

\usepackage{todonotes}

\usepackage[noend]{algpseudocode}   

\newcommand{\frag}{fragile complexity\xspace}

\newcommand{\f}[1]{\ensuremath{f({#1})}}

\newcommand{\Oh}{\ensuremath{\mathcal{O}}}

\let\emptyset\varnothing

\ifComments
\newcommand{\peyman}[1]{{\color{purple}\#\#{(P:)\footnotesize{ #1 }}\#\#}}
\newcommand{\irina}[1]{{\color{purple}\#\#{(I:)\footnotesize{ #1 }}\#\#}}
\newcommand{\manuel}[1]{{\color{purple}\#\#{(M:)\footnotesize{ #1 }}\#\#}}
\newcommand{\david}[1]{{\color{purple}\#\#{(D:)\footnotesize{ #1 }}\#\#}}
\newcommand{\riko}[1]{{\color{purple}\#\#{(R:)\footnotesize{ #1 }}\#\#}}
\newcommand{\nodari}[1]{{\color{purple}\#\#{(N:)\footnotesize{ #1 }}\#\#}}
\newcommand{\uli}[1]{{\color{purple}\#\#{(U:)\footnotesize{ #1 }}\#\#}}
\newcommand{\john}[1]{{\color{purple}\#\#{(J:)\footnotesize{ #1 }}\#\#}}
\newcommand{\rolf}[1]{{\color{purple}\#\#{(Rolf:)\footnotesize{ #1 }}\#\#}}
\newcommand{\pilar}[1]{{\color{purple}\#\#{(Pilar:)\footnotesize{ #1 }}\#\#}}
\newcommand{\stefan}[1]{{\color{purple}\#\#{(Stefan:)\footnotesize{ #1 }}\#\#}}
\else
\newcommand{\peyman}[1]{}
\newcommand{\irina}[1]{}
\newcommand{\manuel}[1]{}
\newcommand{\david}[1]{}
\newcommand{\riko}[1]{}
\newcommand{\nodari}[1]{}
\newcommand{\uli}[1]{}
\newcommand{\john}[1]{}
\newcommand{\rolf}[1]{}
\newcommand{\pilar}[1]{}
\newcommand{\stefan}[1]{}
\fi

\newcommand{\ignore}[1]{}

\let\orgSubSect\subsection
\def\subsection#1{\vspace{-1.1ex}\orgSubSect{#1}\vspace{-.5ex}}

\newcommand{\INV}{\ensuremath{\mathrm{Inv}}}
\newcommand{\RUNS}{\ensuremath{\mathrm{Runs}}}

\begin{document}
\title{Fragile Complexity of Adaptive Algorithms}
%
%
\author{Prosenjit Bose\inst{1}\orcidID{0000-0002-1825-0097} \and
Pilar Cano\inst{2}\orcidID{0000-0002-4318-5282} \and
Rolf Fagerberg\inst{3}\orcidID{0000-0003-1004-3314}\and
John Iacono\inst{2,4}\orcidID{0000-0001-8885-8172}\and
Riko Jacob\inst{5}\orcidID{0000-0001-9470-1809}\and
Stefan Langerman\inst{2}\orcidID{0000-0001-6999-3088}}
\authorrunning{P. Bose, P. Cano, R. Fagerberg, J. Iacono,  R. Jacob, and S. Langerman}
%
\institute{School of Computer Science, Carleton University, Canada. \email{jit@scs.carleton.ca} \and
Université libre de Bruxelles, Belgium.
\email{\{pilar.cano, jiacono, stefan.langerman\}@ulb.ac.be}\\
\and
University of Southern Denmark, Denmark. \email{rolf@imada.sdu.dk}\\
\and
New York University, USA\\
\and
IT University of Copenhagen, Denmark.
\email{rikj@itu.dk}}
\maketitle              
\begin{abstract}
  The \frag of a comparison-based algorithm is $f(n)$ if each input
  element participates in $O(f(n))$ comparisons.
  In this paper, we explore the \frag of algorithms adaptive to
  various restrictions on the input, i.e., algorithms with a \frag
  parameterized by a quantity other than the input size~$n$.  We show
  that searching for the predecessor in a sorted array has \frag
  $\Theta(\log k)$, where $k$ is the rank of the query element, both
  in a randomized and a deterministic setting. For predecessor
  searches, we also show how to optimally reduce the amortized \frag
  of the elements in the array. We also prove the following results:
  Selecting the $k$th smallest element has expected \frag
  $O(\log\log k)$ for the element selected.  Deterministically finding
  the minimum element has \frag $\Theta(\log(\INV))$ and
  $\Theta(\log(\RUNS))$, where $\INV$ is the number of inversions in a
  sequence and $\RUNS$ is the number of increasing runs in a sequence.
  Deterministically finding the median has \frag
  $O(\log(\RUNS) + \log\log n)$ and $\Theta(\log (\INV))$.
  Deterministic sorting has \frag $\Theta(\log (\INV))$ but it has
  \frag $\Theta(\log n)$ regardless of the number of runs.
 
 \keywords{Algorithms \and Comparison based algorithms \and Fragile complexity.}
\end{abstract}

\input{intro}

\input{search}

\input{selection2}

\input{sorting}

\paragraph{Acknowledgements.} 
\begin{small}
This material is based upon work performed while attending AlgoPARC Workshop on Parallel Algorithms and Data Structures at the University of Hawaii at Manoa, in part supported by the National Science Foundation under Grant No.~CCF-1930579. We thank Timothy Chan and Qizheng He for their ideas improving the randomized selection algorithm. P.B was partially supported by NSERC. P.C and J.I. were supported by F.R.S.-FNRS under Grant no MISU F 6001 1. R.F. was partially supported by the Independent Research Fund Denmark, Natural Sciences, grant DFF-7014-00041. J.I. was supported by NSF grant CCF-1533564. S.L. is Directeur de Recherches du F.R.S.-FNRS.
\end{small}

\bibliographystyle{splncs04}
\bibliography{CIAC-camera_ready.bbl}

\end{document}

%% file: intro.tex
\section{Introduction}
\label{sec:intro}

Comparison-based algorithms have been thoroughly studied in computer
science. This includes algorithms for problems such as
\textsc{Minimum}, \textsc{Median}, \textsc{Sorting},
\textsc{Searching}, \textsc{Dictionaries}, \textsc{Priority Queues},
and many others. The cost measure analyzed is almost always the total
number of comparisons performed by the algorithm, either in the worst
case or the expected case. Recently, another type of cost measure has
been introduced~\cite{afshani:fragile-ESA19} which instead considers
how many comparisons each individual element is subjected during the
course of the algorithm. In~\cite{afshani:fragile-ESA19}, a
comparison-based algorithm is defined to have {\em \frag} $\f{n}$ if
each individual input element participates in at most $\f{n}$
comparisons. The \frag of a computational problem is the best possible
\frag of any comparison-based algorithm solving the problem.
\stefan{Define out-\frag complexity}

This cost measure has both theoretical and practical motivations. On
the theoretical side, it raises the question of to what extent the
comparisons necessary to solve a given problem can be spread evenly
across the input elements. On the practical side, this question is
relevant in any real world situation where comparisons involve some
amount of destructive impact on the elements being compared (hence the
name of the cost measure). As argued in~\cite{afshani:fragile-ESA19},
one example of such a situation is ranking of any type of consumable
objects (wine, beer, food, produce), where each comparison reduces the
available amount of the objects compared. Here, an algorithm like
\textsc{QuickSort}, which takes a single object and partitions the
whole set with it, may use up this pivot element long before the
algorithm completes. Another example is sports, where each comparison
constitutes a match and takes a physical toll on the athletes
involved. If a comparison scheme subjects one contestant to many more
matches than others, both fairness to contestants and quality of
result are impacted---finding a winner may not be very useful if this
winner has a high risk of being injured in the process. The negative
impact of comparisons may also be of non-physical nature, for instance
when there is a privacy risk for the elements compared, or when bias
grows if few elements are used extensively in comparisons.



\subsection{Previous work}
In~\cite{afshani:fragile-ESA19}, the study of algorithms' \frag was
initiated and a number of upper and lower bounds on the \frag for
fundamental problems was given. The problems studied included
\textsc{Minimum}, the \textsc{Selection}, \textsc{Sorting}, and
\textsc{Heap Construction}, and both deterministic and randomized
settings were considered.
In the deterministic setting, \textsc{Minimum} was shown to have \frag
$\Omega(\log n)$ and \textsc{Sorting} to have \frag $O(\log n)$. Since
\textsc{Sorting} can solve \textsc{Selection}, which can solve
\textsc{Minimum}, the \frag of all three problems is $\Theta(\log
n)$. The authors then consider randomized algorithms, as well as a
more fine-grained notion of \frag, where the objective is to protect
selected elements such as the minimum or median (i.e., the element to
be returned by the algorithm), possibly at the expense of the
remaining elements.
Among other results, it is shown in~\cite{afshani:fragile-ESA19} that
\textsc{Minimum} can be solved incurring expected $O(1)$ comparisons
on the minimum element itself, at a price of incurring expected
$O(n^{\varepsilon})$ on each of the rest. Also a more general
trade-off between the two costs is shown, as well as a close to
matching lower bound. For \textsc{Selection}, similar results are
given, including an algorithm incurring expected $O(\log \log n)$
comparisons on the returned element itself, at a price of incurring
expected $O(\sqrt{n})$ on each of the rest.

An earlier body of work relevant for the concept of \frag is the study
of sorting networks, started in 1968 by Batcher~\cite{Batc68}. In
sorting networks, and more generally comparator networks, the notion
of depth (the number of layers, where each layer consists of
non-overlapping comparators) and size (the total number of
comparators) correspond to fragile complexity and standard worst case
complexity, respectively, in the sense that a network with
depth~$f(n)$ and size~$s(n)$ can be converted into a comparison-based
algorithm with fragile complexity~$f(n)$ and standard
complexity~$s(n)$ by simply simulating the network.

Batcher, as well as a number of later
authors~\cite{Dowd-Perl-Rudolph-Saks/89,Parberry/92,IPL::ParkerP1989,books/garland/Pratt72}, gave sorting networks with
$\Oh(\log^2 n)$ depth and $\Oh(n\log^2 n)$ size.
For a long time it was an open question whether better results were
possible. In 1983, Ajtai, Koml{\'o}s, and
Szemer{\'e}di~\cite{aks-halvers,aks} answered this in the affirmative
by constructing a sorting network of $\Oh(\log n)$ depth and
$\Oh(n\log n)$ size. This construction is quite complex and involves
expander graphs~\cite{HooLinWig06,journals/fttcs/Vadhan12}. It was
later modified by
others~\cite{Chvatal/92,conf/stoc/Goodrich14,Paterson/90,journals/algorithmica/Seiferas09},
but finding a simple, optimal sorting network, in particular one not
based on expander graphs, remains an open problem.  Comparator
networks for other problems, such as selection and heap construction
have also been studied
~\cite{alekseev:selection-69,brodal:heap-98,Jimbo,Pippenger91,Yao.merging}.

While comparator networks are related to \frag in the sense that
results for comparator networks can be transfered to the \frag setting by
simple simulation, it is demonstrated in~\cite{afshani:fragile-ESA19}
that the two models are not equivalent: there are problems where one
can construct fragile algorithms with the same \frag, but with
strictly lower standard complexity (i.e., total number of comparisons)
than what is possible by simulation of comparison networks. These
problems include \textsc{Selection} and \textsc{Heap Construction}.

\subsection{Our Contribution}
In many settings, the classical worst case complexity of
comparison-based algorithms can be lowered if additional information
on the input is known. For instance, sorting becomes easier than
$\Theta(n \log n)$ if the input is known to be close to
sorted. Another example is searching in a sorted set of elements,
which becomes easier than $O(\log n)$ if we know an element of rank
close to the element searched for. Such algorithms may be described as
\emph{adaptive} to input restrictions (using the terminology from the
sorting setting~\cite{DBLP:journals/csur/Estivill-CastroW92}). Given
that the total number of comparisons can be lowered in such
situations, the question arises whether also reductions in the \frag
are possible under these types of input restrictions.

In this paper, we expand the study of the \frag of comparison-based
algorithms to consider the impact of a number of classic input
restrictions. We show that searching for the predecessor in a sorted
array has \frag $\Theta(\log k)$, where $k$ is the rank of the query
element, both in a randomized and a deterministic setting. For
predecessor searches, we also show how to optimally reduce the
amortized \frag of the elements in the array. We also prove the
following results: Selecting the $k$th smallest element has expected
\frag $O(\log\log k)$ for the element selected.  Deterministically
finding the minimum element has \frag $\Theta(\log(\INV))$ and
$\Theta(\log(\RUNS))$, where $\INV$ is the number of inversions in a
sequence and $\RUNS$ is the number of increasing runs in a sequence.
Deterministically finding the median has \frag
$O(\log(\RUNS) + \log\log n)$ and $\Theta(\log (\INV))$.
Deterministic sorting has \frag $\Theta(\log (\INV))$ but it has \frag
$\Theta(\log n)$ regardless of the number of runs.

%% file: search.tex
\newcommand{\lrp}[1]{\left( #1 \right)}
\newcommand{\lrc}[1]{\lceil #1 \rceil}

\section{Searching}
\label{sec:searching}

The problem of predecessor searching is, given a sorted array~$A$ with
$n$ elements, $A[0]..A[n-1]$, answer queries of the form ``What is the
index of the largest element in $A$ smaller than $x$?''  Binary search
is the classic solution to the predecessor search problem. It achieves
$\log n$ \frag for~$x$, and \frag at most one for each element of~$A$.
We can improve on this in two ways.  The first is where we try to keep
the \frag of~$x$ small, which is possible if we know something about
the rank of~$x$. We show that the optimal dependency on
the rank of~$x$ is $\Theta(\log k)$ where $k$ is its rank, both for deterministic and
randomized algorithms.\footnote{For simplicity of exposition, we assume the
  rank is close to one, but the result clearly holds for rank distance
  to other positions in~$A$.}
The second setting is where we are concerned with the
\frag of the other elements.  While there is no way to improve a
single search, classical deterministic binary search will always do
the first comparison with the same element (typically the median).
Hence we consider deterministic algorithms that improve the amortized
\frag of any element of the array~$A$ over a sequence of searches.

\subsection{Single search}

\begin{theorem}\label{thm:rank}
  Let $A$ be a sorted array.
  Determining the predecessor of an element $x$ within~$A$ has \frag $\Theta(\log k)$ for deterministic and randomized algorithms, where $k$ is the rank of $x$ in~$A$. 
\end{theorem}
\begin{proof}
The upper bound follows from standard exponential
search~\cite{conf/stoc/Fredman75}: We compare~$x$ to
$A[2],A[4],A[8],\ldots$ until we find the smallest~$i$ such that
$x<A[2^i]$.  We perform a binary search with the initial interval
$[2^{i-1},2^i]$.  If $x$ has the predecessor $A[k]$, this requires
$O(\log k)$ comparisons.


For the lower bound assume we have a deterministic algorithm to
determine the rank of an element~$x$.  If the answer of the algorithm
is~$k$, let $B_k$ be the bit-string resulting from concatenating the
sequence of the outcomes of the comparisons performed by the
algorithm, the $i$-th bit $B_k[i]=0$ for $x<A[k]$, otherwise it is~1.
Because the algorithm is deterministic and correct, all these
bit-strings are different and they are a code for the numbers
$1,\ldots,n$.  Now, for any~$k$, consider the uniform distribution on
the numbers~$0,\ldots,k-1$, a distribution with entropy~$\log k$.  By
Shannon's source coding theorem, the average code length must be at
least $\log k$, i.e., $\sum_{i=0}^{k-1} |B_i| \ge k\log k$.

For a contradiction, assume there would be an algorithm with $|B_i|\le \log i$ (the binary logarithm itself). 
Then for $k> 1$, $\sum_{i=0}^{k-1} |B_i| < k\log k$, in contrast to Shannon's theorem.

The bound $\sum_{i=0}^{k-1} |B_i| \ge k\log k$ also holds for randomized algorithms if the queries are drawn uniformly from~$[1,\ldots,k]$, following Yao's principle:
Any randomized algorithm can be understood as a collection of deterministic algorithms from which the 'real' algorithm is drawn according to some distribution. 
Now each deterministic algorithm has the lower bound, and the average number of comparisons of the randomized algorithm is a weighted average of these. 
Hence the lower bound also holds for randomized algorithms.\qed 
\end{proof}
\subsection{Sequence of searches}

As mentioned, in binary search, the median element of the array will be compared with every query element. Our goal here is to develop a search strategy so as to ensure that data far away from the query will only infrequently be involved in a comparison.
Data close to the query must be queried more frequently. While we prove this formally in Theorem~\ref{t:lb}, it is easy to see that predecessor and successor of a query must be involved in comparisons with the query in order to answer the query correctly. 

\begin{theorem} \label{t:search}
There is a search algorithm that 
for any sequence of predecessor searches $x_1, x_2, \ldots, x_m$ in a sorted array $A$ of size $n$
the number of comparisons with any $y\in A$ is $O\lrp{ \log n+ \sum_{i=1}^m \frac{1}{d(x_i,y)}}$ 
 where $d(x,y)$ is the number of elements between $x$ and $y$ in $A$, inclusive. 
The runtime is $O(\log n)$ per search and the structure uses $O(n)$ bits of additional space.
\end{theorem}

\def\offset{\text{\sl offset}}
\begin{proof}
We use the word interval to refer to a contiguous range of $A$; when we index an interval, we are indexing $A$ relative to the start of the interval.
Call an aligned interval $I$ of $A$ of rank $i$ to be $( A[k\cdot 2^i]\ldots A[(k+1) \cdot 2^i])$ for some integer $k$, i.e., the aligned intervals of $A$ are the dyadic intervals of $A$. There are $O(n)$ aligned intervals of $A$, and for each aligned interval $I$ of rank $i$ we store an offset $I.\offset$ which is in the range $[0,2^i)$, and it is initialized to 0.

The predecessor search algorithm with query $x$ is a variant of recursive binary search, where at each step an interval $I_q$ of $A$ is under consideration, and the initial recursive call considers the whole array $A$. Each recursive call proceeds as follows: Find the largest $i$ such that there are at least three rank-$i$ aligned intervals in $I_q$, use $I_m$ to denote  the middle such interval (or an arbitrary non-extreme one if there are more than three), and we henceforth refer to this recursive call as a rank-$i$ recursion. Compare $I_m[I_m.\offset]$ with $x$, and then increment $I_m.\offset$ modulo $2^i$. Based on the result of the comparison, proceed recursively as in binary search.
The intuition is by moving the offset with every comparison, this prevents a single element far from the search from being accessed too frequently.
We note that the total space used by the offsets is $O(n)$ words, which can be reduced to $O(n)$ bits if the offsets are stored in a compact representation.

\pagebreak[3]
\noindent First, several observations:\nopagebreak
\begin{compactenum}
\item \label{p1} In a rank-$i$ recursion, $I_q$ has size at least $3 \cdot 2^i$ (since there must be at least three rank-$i$, size $2^i$ aligned intervals in $I_q$) and at most $8 \cdot 2^i$, the latter being true as if it was this size there would be three rank-$i+1$ intervals in $I_q$, which would contradict $I_m$ having rank $i$. 

\item If $I_q$ has size $k$ then if there is a recursive call, it is called with an interval of size at most~$\frac{7}{8}k$. This is true by virtue of $I_m$ being rank-$i$ aligned with at least one rank-$i$ aligned interval on either side of $I_m$ in $i$. Since $I_q$ has size at most $8\cdot 2^i$, this guarantees an eighth of the elements of $I_q$ will be removed from consideration as a result of the comparison in any recursive call.

\item \label{p3} From the previous two points, one can conclude that for a given rank $i$, during any search there are at most 7 recursions with of rank $i$. This is because after eight recursions any rank-$i$ search will be reduced below the minimum for rank $i$:  $8 \cdot 2^i \cdot \lrp{\frac{7}{8}}^8 <3 \cdot 2^i$.
\end{compactenum}

For the analysis, we fix an arbitrary element $y$ in $A$ and use the potential method to analyse the comparisons involving $y$. 
Let $\mathcal{I}_y=\{I_y^1,I_y^2\ldots\}$ be the $O(\log n)$ aligned intervals that contain $y$, numbered such that $I_y^i$ has rank $i$.
Element $y$ will be assigned a potential relative to each aligned interval $I_y^i\in \mathcal{I}_y$ which we will denote as $\varphi_y(I_y^i)$. 
Let $t_y(I_y^i)$ be number of times $I_y^i.\offset$ needs to be incremented before $I_y^i[I_y^i.\offset]=y$, which is in the range $[0,2^i)$. 
The potential relative to $I^i_y$ is then defined as $\varphi_y(I_y^i)\coloneqq \frac{2^i-t_y(I^i_y)}{2^i}$, and the potential relative to $y$ is defined to be the sum of the potentials relative to the intervals in $\mathcal{I}_y$: $\varphi_y \coloneqq \sum_{I^i_y\in\mathcal{I}_y} \varphi_y(I_y^i)$.

How does $\varphi_y(I_y^i)$ change during a search? 
First, if there is no rank-$i$ recursive call during the search to an interval containing $y$, it does not change as $I_y^i.\offset$ is unchanged.
Second, observe from point \ref{p3} that a search can increase $\varphi_y (I_y^i)$ by only $\frac{7}{2^i}$.
Furthermore if $y$ was involved in a comparison during a rank-$i$ recursion, there will be a loss of $1-\frac{1}{2^i}$ units of potential in $\varphi_y(I_y^i)$ as the offset of $I_y^i$ changes from 0 to $2^i-1$.

Following standard potential-based amortized analysis, the amortized number of comparisons involving $y$ during a search is the actual number of comparisons (zero or one) plus the change in the potential $\varphi_y $. 
Let $i_{\min}$ be the smallest value of $i$ for which there was a rank-$i$ recursion that included $y$. As the maximum gain telescopes, the potential gain is at most $\frac{14}{2^{i_{\min}}}$, minus 1 if $y$ was involved in a comparison. 
Thus the amortized number of comparisons with $y$ in the search is at most $\frac{14}{2^{i_{\min}}}$.

Observe that if there was a rank-$i$ recursion that included $y$, that $d(x,y)$ is at most $8\cdot 2^i$ by point~\ref{p1}.
This gives $d(x,y)\leq 8\cdot 2^i \leq 8\cdot 2^{i_{\min}}$.
Thus the amortized cost  can be restated as being at most
$ \frac{14}{2^{i_{\min}}}\leq \frac{112}{d(x,y)}$. 

To complete the proof, the total number of comparisons involving $y$ over a sequence of searches is the sum of the amortized costs plus any potential loss. As the potential $\varphi_y$ is always nonnegative and at most $\lrc{\log n}$ (1 for each $\varphi_y(I^i_y)$), this gives the total cost as $O\lrp{ \log n+ \sum_{i=1}^m \frac{1}{d(x_i,y)}}$. \qed
\end{proof}

Note that the above proof was designed for easy presentation and not an optimal constant. Also note that this theorem implies that if the sequence of searches is uniformly random, the expected fragility of all elements is $O(\frac{\log n}{n})$, which is asymptotically the best possible since random searches require $\Omega(\log n)$ comparisons in expectation.

\subsection{Lower Bounds.} It is well-known that comparison-based searching requires $\Omega(\log n)$ comparisons per search. In our method, taking a single search $x_i$ summing over the upper bound on amortized cost of the number of comparisons with $y$, $\frac{42}{d(x_i,y)}$, for all $y$ yields a harmonic series which sums to $O(\log n)$. But we can prove something stronger:

\begin{theorem} \label{t:lb}
There is a constant $c$ such that if a predecessor search algorithm has an amortized number of comparisons of $f(d(x_i,y))$ for an arbitrary $y$ for every sequence of predecessor searches $x_1, x_2, \ldots x_m$, then $\sum_{k=1}^p f(k) \geq c \log p$ for all $p \leq n$.
\end{theorem}

\begin{proof}
This can be seen by looking at a random sequence of predecessor searches for which the answers are uniform among $A[0]\ldots A[p-1]$, if the theorem was false, similarly to the proof of Theorem~\ref{thm:rank}, this would imply the ability to execute such a sequence in $o(\log p)$ amortized time per operation. \qed
\end{proof}
\riko{perhaps we could do this even nicer}

This shows that a flatter asymptotic tradeoff between $d(x_i,y)$ and the amortized comparison cost is impossible; more comparisons are needed in the vicinity of the search than farther away.
For example, a flat amortized number of comparisons of $\frac{\log n}{n}$ for all elements would sum up to $O(\log n)$ amortized comparisons over all elements, but yet would violate this theorem. 

\subsection{Extensions.}

Here we discuss extensions to the search method above. We omit the proofs as they are simply more tedious variants of the above.

One can save the additional space used by the offsets of the intervals through the use of randomization. The offsets force each item in the interval to take its turn as the one to be compared with, instead one can pick an item at random from the interval. This can be further simplified into a binary search where at each step one simply picks a random element for the comparison amongst those (in the middle half) of the part of the array under consideration. 

To allow for insertions and deletions, two approaches are possible. The first is to keep the same array-centric view and simply use the packed-memory array \cite{DBLP:conf/icalp/ItaiKR81,DBLP:conf/sigmod/Willard86,DBLP:journals/iandc/Willard92} to maintain the items in sorted order in the array. This will give rise to a cost of $O(\log^2 n)$ time which is inherent in maintaining a dynamic collection of items ordered in an array \cite{DBLP:journals/siamcomp/BulanekKS15} (but no additional fragility beyond searching for the item to insert or delete as these are structural changes).
The second approach would be to use a balanced search tree such as a red-black tree \cite{DBLP:conf/focs/GuibasS78}. This will reduce the insertion/deletion cost to $O(\log n)$ but will cause the search cost to increase to $O(\log^2 n)$ as it will take $O(\log n)$ time to move to the item in each interval indicated by the offset, or to randomly choose an item in an interval. The intervals themselves would need to allow insertions and deletions, and would, in effect be defined by the subtrees of the red-back tree.
It remains open whether there is a dynamic structure with the fragility results of Theorem~\ref{t:search} where insertions and deletions can be done in $O(\log n)$ time.

%% file: selection2.tex
\section{Selection}
\label{sec:selection}



In this section we consider the problem of finding the $k$-th smallest element of an unsorted array.
There is a randomized algorithm that selects the $k$-th smallest element with expected fragile complexity of $O(\log \log n)$ for the selected element~\cite{afshani:fragile-ESA19}.
We consider the question if this complexity can be improved for small $k$.
In this section we define a sampling method that, combined with the algorithm given in~\cite{afshani:fragile-ESA19}, selects the $k$-th smallest element with expected $O(\log \log k)$ comparisons.

Next, we define the filtering method \textsc{ReSet} in a tail-recursive fashion. The idea of this procedure is the following: First, we build a random half size sample $A_1$ from the input set $X$. Later, we continue recursively constructing a random half sample $A_i$ from the previous sample $A_{i-1}$ until we get a random sample $A_{\ell}$ of size $O(k+1)$. Once $A_{\ell}$ is given, then a set $A'_{\ell}$ of size $O(k)$ is given for the previous recursive call. Using such set, a new subset $A'_{\ell-1}$ is given from the previous sample $A_{\ell-1}$ where its expected size is $O(k)$. This process continuous until a final subset $\mathcal{C}$ is given from the input set $X$ such that its expected size $O(k)$ and it contains the $k$-th smallest element of $X$. 

\noindent\fbox{\scalebox{0.95}{\begin{minipage}{\textwidth}
\begin{small}
\begin{algorithmic}[1]
	 \Procedure{ReSet}{$X, k$}
	 	\Comment{Returns a small subset $\mathcal{C}$ of $X$ that contains the $k$-th element}
			\State Let $n=|X|$ and $\mathcal{C} = \emptyset$
			\If{$k \geq \frac{n}{2}-1$}
			    \Comment{The set has size $O(k+1)$}
			    \State Let $A'=X$
			\Else
			    \Comment{Recursively construct a sample of expected size $O(k+1)$}
			    \State Sample $A$ uniformly at random from $X$, $|A|= \frac{n}{2}$
			    \State Let $A'=$ \textsc{ReSet($A, k$)}
			\EndIf
			\State Choose the ($k+1$)-th smallest element $z$ from $A'$ (by standard linear time selection) 
			\State Let $\mathcal{C}=\{x \in X : x\leq z\}$
			\State \Return $\mathcal{C}$
	\EndProcedure
\end{algorithmic}
\end{small}
\end{minipage}}
}

In the following theorem we show that the combination of the \textsc{ReSet} procedure and the randomized selection algorithm in~\cite{afshani:fragile-ESA19}, results in expected $O(\log \log k)$ comparisons for the selected element. 

\begin{theorem}
Randomized selection is possible in expected fragile complexity\\ $O(\log \log k)$ in the selected element.
\end{theorem}
\begin{proof}
Let us show that the following procedure for selecting the $k$-th element in a set $X$ with $|X|=n$, gives an expected fragile complexity $O(\log \log k)$ in the $k$-th element:

\emph{If $k > n^{\frac{1}{100}}$, then let $S'=X$. If $k \leq n^{\frac{1}{100}}$, then sample uniformly at random $S$ from $X$, where $|S|=\frac{n}{k}$. Let $C=$ \textsc{ReSet}($S, k$) and select the $k+1$-th smallest element $z$ from $\mathcal{C}$ by standard linear time selection. Let $S'=\{x \in X: x\leq z\}$. Finally, apply to $S'$ the randomized selection algorithm of~\cite{afshani:fragile-ESA19}.}

Let $x_k$ denote the $k$-th smallest element in $X$ and let $f_k$ denote the \frag of $x_k$. 
Note that if $x_k \in S$, then, before constructing $S'$, $f_k$ is given by the \frag of $x_k$ in $\textsc{ReSet}(S,k)$ plus $O(|\mathcal{C}|)$ when finding the ($k+1$)-th smallest element in $\mathcal{C}$. Otherwise, $x_k$ is not compared until $S'$ is constructed. 
On the other hand, recall that the expected $f_k$ in the algorithm in~\cite{afshani:fragile-ESA19} is $O(\log \log m)$ where $m$ is the size of the input set. Hence, the expected $f_k$ after selecting the $k+1$-th element in  $\mathcal{C}$ is 1 when creating $S'$ plus the expected $f_k$ in the randomized selection algorithm in~\cite{afshani:fragile-ESA19} that is $\sum_{|S'|}O(\log \log |S'|)\mathbb{P}[|S'|]=\mathbb{E}[O(\log \log |S'|)]$. Thus, $\mathbb{E}[f_k]=(\mathbb{E}[f_k\text{ in \textsc{ReSet}}| x_k \in S]+\mathbb{E}[|\mathcal{C}|])\mathbb{P}[x_k \in S] + 0\mathbb{P}[x_k \notin S] + 1 + \mathbb{E}[O(\log \log |S'|)]$. Since the logarithm is a concave function,  $\mathbb{E}[O(\log \log |S'|)]\leq O(\log \log (\mathbb{E}[|S'|]))$.  
Therefore, if we prove that: (i) the expected \frag of $x_k$ before creating $S'$ is $O(1)$ and (ii) $\mathbb{E}[|S'|]=c'k^{c}$ for some constants $c$ and $c'$. Then, we obtain that $\mathbb{E}[f_k]\leq O(1)+1+O(c\log\log k+\log c') = O(\log \log k)$, as desired. In order to prove (i) and (ii) we consider 2 cases: (1) $k > n^{\frac{1}{100}}$, (2) $k \leq n^{\frac{1}{100}}$.
\\
\emph{Case 1)} $S' = X$ and it makes no previous comparisons in any element, proving (i). In addition, $S'$ has size less than $k^{100}$. Thus, (ii) holds. 
\\
\emph{Case 2)} $S$ is a sample of $X$ with size $\frac{n}{k}$ and $S'=\textsc{ReSet}(S, k)$. \\
First, let us show (i). If $x_k \notin S$, then there are no previous comparisons. Hence, the expected \frag of $x_k$ before constructing $S'$ is given by $(\mathbb{E}[f_k\text{ in \textsc{ReSet}}| x_k \in S'] + \mathbb{E}[|\mathcal{C}|])\mathbb{P}[x_k \in S]+ 0$. Since $S$ is an uniform random sample with size $\frac{n}{k}$, $\mathbb{P}[x_k \in S] = \frac{1}{k}$, it suffices to show that $\mathbb{E}[f_k\text{ in \textsc{ReSet}}| x_k \in S'] + \mathbb{E}[|\mathcal{C}|]=O(k)$, which gives an expectation of $O(k)\frac{1}{k}=O(1)$, proving (i). So, let us show that $\mathbb{E}[f_k\text{ in \textsc{ReSet}}| x_k \in S'] + \mathbb{E}[|\mathcal{C}|]=O(k)$. 
Let $A_0=S$ and let $A_1$ be the sample of $A_0$ when passing through line 6 in \textsc{ReSet}. Similarly, denote by $A_i$ to the sample of $A_{i-1}$ in the $i$-th recursive call of \textsc{ReSet} and let $A'_i=\textsc{ReSet}(A_{i}, k)$. Note that by definition $A'_0=\mathcal{C}$. 
Let $\ell+1$ be the number of recursive calls in $\textsc{ReSet}(S, k)$.  

Since $A_i$ is a uniform random sample of size $\tfrac{|A_{i-1}|}{2}$ for all $i\geq 1$, $\mathbb{P}[x \in A_i| x \in A_{i-1}]=2^{-1}$ and $\mathbb{P}[x \in A_i| x \notin A_{i-1}]=0$. Hence, $\mathbb{P}[x_{k} \in A_i]=\mathbb{P}[x_k \in \cap_{i=0}^{i} A_i]=2^{-i}$. Note that the number of comparisons of $x_k$ in \textsc{ReSet} is given by the number of times $x_k$ is compared in lines 8 and 9. Thus, for each $i$-th recursive call: if $x_k \in A_i$, then $x_k$ is compared once in line 9; and if $x_k \in A_i\cap A'_i$, then $x_k$ is compared at most $|A'_i|$ times in line 8. Otherwise, $x_k$ is not compared in that and the next iterations.  
Thus, $\mathbb{E}[f_k\text{ in \textsc{ReSet}}| x_k \in S'] + \mathbb{E}[|\mathcal{C}|] \leq \sum_{i=0}^{\ell} (1+\mathbb{E}[|A'_i|])\mathbb{P}[x_k \in A_i]= \sum_{i=0}^{\ell} 2^{-i}(1+\mathbb{E}[|A'_i|])\leq 2(1+ \mathbb{E}[|A'_i|])$. 
Let us compute $\mathbb{E}[|A'_i|]$.  
Since the ($\ell$$+$$1$)-th iteration $\textsc{ReSet}(A_{\ell},k)$ passes through the if in line 3, there is no new sample from $A_{\ell}$. 
Thus, $A'_{\ell}$ is given by the $k+1$ smallest elements of $A_{\ell}$. Therefore, $\mathbb{E}[|A'_{\ell}|]=k+1$ Denote by $a'^{i}_j$ to the $j$-th smallest element of $A'_i$. For the case of $0\leq i<\ell$, we have $A'_i=\{x\in A_{i+1} : x\leq a'^{i+1}_{k+1}\}$. Hence, $\mathbb{E}[|A'_i|]=\mathbb{E}[|\{x\in A_{i+1} : x\leq a'^{i+1}_{k+1}\}|] =\mathbb{E}[|\{x \in A_{i+1} : x \leq a'^{i+1}_1]\}|] + \sum_{j=1}^{k} \mathbb{E}[|\{x \in A_{i+1} :  a'^{i+1}_{j-1}< x \leq a'^{i+1}_{j}]\}|]\leq \sum_{j=1}^{k+1}\sum_{t=1}^{\infty} t2^{-1}(2^{-(t-1)})=2(k+1)$. 
Therefore, $\mathbb{E}[f_k\text{ in \textsc{ReSet}}| x_k \in S'] + \mathbb{E}[|\mathcal{C}|]=\sum_{i=0}^{\ell} 2^{-i}(1+\mathbb{E}[|A'_i|])\leq 2+ 2\mathbb{E}[|A'_i|]=O(k)$ proving (i). 
Finally, let us show (ii): For simplicity, let $c_j$ denote the $j$-th smallest element of $\mathcal{C}$. Then, $\mathbb{E}[|S'|]=\mathbb{E}[|\{x\in X: x\leq c_1\}|]+\sum_{j=1}^{k}\mathbb{E}[|\{x\in X: c_j \leq x\leq c_{j+1}\}|]\leq \sum_{j=1}^{k+1}\sum_{j=0}^{\infty} j k^{-1}(1-k^{-1})^{j-1}=k(k+1)=O(k^2)$, proving (ii). \qed
\end{proof}

%% file: sorting.tex
\section{Sorting}
\label{sec:sorting}

When the input is known to have some amount of existing order, sorting
can be done faster than $\Theta(n \log n)$. Quantifying the amount of
existing order is traditionally done using measures of
disorder~\cite{DBLP:journals/csur/Estivill-CastroW92}, of which \INV{}
and \RUNS{} are two classic examples.\footnote{The measure \INV{} is
  defined as the total number of inversions in the input, where each
  of the ${n\choose 2}$ pairs of elements constitute an inversion if
  the elements of the pair appear in the wrong order. The measure
  \RUNS{} is defined as the number of runs in the input, where a run is a maximal consecutive ascending subsequence.}
A sorting algorithm is adaptive to a measure of disorder if it
is faster for inputs with a smaller value of the measure. For the
above measures, run times of $O(n \log (\INV/n))$ and
$O(n \log (\RUNS))$ can be achieved. These results are best possible
for comparison-based sorting, by standard information-theoretic
arguments based on the number of different inputs having a given maximal
value of the measure.

The fact~\cite{aks,afshani:fragile-ESA19} that we can sort all inputs
in $\Theta(n \log n)$ time and $\Theta(\log n)$ \frag can be
viewed as being able to distribute the necessary comparisons evenly
among the elements such that each element takes part in at most
$\Theta(\log n)$ comparisons.
%
Given the running times for adaptive sorting stated above, it is
natural to ask if for an input with a given value of \INV{} or
\RUNS{} we are able to sort in a way that distributes the necessary
comparisons evenly among the elements, i.e., in a way such that each
element takes part in at most $O(\log (\INV))$ or $O(\log (\RUNS))$
comparisons, respectively.
In short, can we sort in \frag $O(\log (\INV))$ and
$O(\log (\RUNS))$?  Or more generally, what problems can we solve with
\frag adaptive to \INV{} and \RUNS{}? In this section, we study
the \frag of deterministic algorithms for \textsc{Minimum},
\textsc{Median}, and \textsc{Sorting} and essentially resolve their
adaptivity to \INV{} and \RUNS{}.

\begin{theorem}
\textsc{Minimum} has \frag $\Theta(\log (\RUNS))$.
\end{theorem}
\begin{proof}
For the upper bound: identify the runs in $O(1)$ \frag by a scan of the input. Then, use a
tournament on the heads of the runs since the minimum is the minimum of the heads of the runs. 
For the lower bound: apply the logarithmic lower
bound for \textsc{Minimum}~\cite{afshani:fragile-ESA19} on the heads
of the runs. \qed
\end{proof}

\begin{theorem}
\textsc{Sorting} has \frag $\Theta(\log n)$, no matter what value of
\RUNS{} is assumed for the input.
\end{theorem}
\begin{proof}
The upper bound follows from general sorting. For the lower bound: the input
consisting of a run~$R$ of length $n-1$ and one more element~$x$ has
$\RUNS = 2$, but $\log n$ comparisons on $x$ can be forced by an
adversary before the position of $x$ in $R$ is determined.\qed
\end{proof}

\begin{theorem}
\textsc{Median} has \frag $O(\log (\RUNS) + \log\log n)$.
\end{theorem}
\begin{proof}
Assume that $4 \cdot \RUNS \cdot \log n < n/2$, since otherwise the
claimed \frag is $O(\log n)$ for which we already have a median
algorithm~\cite{afshani:fragile-ESA19}. Consider the rank space
$[1,n]$ (i.e., the indices of the input elements in the total sorted
order) of the input elements and consider the rank interval $[a,b]$
around the median defined by $a = n/2 - 4 \cdot \RUNS \cdot \log n$
and $b = n/2 + 4 \cdot \RUNS \cdot \log n$. In each step of the
algorithm, elements are removed in two ways: type~A removals and
type~B removals. A removal of type~A is a balanced removal, where a
number of elements with ranks in $[1,a-1]$ are removed and the same
number of elements with ranks in $[b+1,n]$ are removed. The key
behind the type~A removal is that the median element of the set
prior to the removal is the same as the median of the set after the
removal, if the median prior to the removal has a rank in $[a,b]$.

A removal of type~B is a removal of elements with arbitrary
rank. However, the total number of elements removed by type~B removals
is at most $7 \cdot \RUNS \cdot \log n$ during the\ entire run of the
algorithm. Hence, repeated use of type~A and type~B removals will
maintain the invariant that the median of the remaining elements has a
rank in $[a,b]$.

We now outline the details of the algorithm. The first step is to
identify all the runs in $O(1)$ \frag by a scan. A run will be
considered {\em short} if the run consists of fewer than
$7 \cdot \log n$ elements and it will be considered {\em long} otherwise. A
step of the algorithm proceeds by first performing a type~B removal
followed by a type~A removal.  A type~B removal consists of removing
all short runs that are present. The short runs that are removed will
be reconsidered again at the end once the number of elements under
consideration by the algorithm is less than
$64 \cdot \RUNS \cdot \log n$.

Once a type~B removal step is completed, only long runs remain under
consideration.  We now describe a type~A removal step. Note that a
long run may become short after a type~A removal step, in which case
it will be removed as part of the next type~B removal step. Each run
can become short (and be removed by a type~B removal) only once, hence
the total number of elements removed by type~B removals will be at
most $7 \cdot \RUNS \cdot \log n$, as claimed.

In the following, let $\mathcal{N}$ denote the elements under
consideration just before a type~A removal (i.e., the elements of the
remaining long runs), and let $N = |\mathcal{N}|$. The algorithm stops
when $N \le 64 \cdot \RUNS \cdot \log n$.

To execute the type~A removal step, the algorithm divides each long
run $R$ into blocks of length $\log n$. The blocks of a run are
partitioned by a {\em partitioning} block. The partitioning block has
the property that there are at least $|R|/7$ elements of $R$ whose
values are less than the values in the partitioning block and at least
$5|R|/7$ elements of $R$ whose value are greater than the elements in
the partitioning block. One element~$x_R$ is selected from the
partitioning block. We will refer to this element as a {\em
partitioning} element.
These partitioning elements are then sorted
into increasing order, which incurs a cost of $O(\log (\RUNS))$ \frag
on each of the partitioning elements.  The runs are then arranged in
the same order as their partitioning elements.  Label this sequence of
runs as $R_1, R_2, \dots, R_k$, and let $t$ be the largest index such
that $\sum_{i=1}^{t-1} |R_i| < N/8.$

Since the partitioning element $x_{R_t}$ is smaller than all the elements in the blocks with values greater than their
respective partitioning blocks
in $R_t, R_{t+1},$ $\dots, R_k$, we have that $x_{R_t}$ is smaller than $(7/8)(5N/7) = 5N/8$ of the
remaining elements. Hence in rank it is at least $N/8$ below the median
of the remaining elements. By the invariant on the position in rank
space of this median and the fact that $N > 64 \cdot \RUNS \cdot \log n$, we note
that $x_{R_t}$ has a rank below~$a$. We also note that all the elements
below the partitioning blocks in $R_1, R_{2}, \dots, R_t$ have value less than 
$x_{R_t}$. This constitutes at least $(1/8)(N/7) = N/56$
elements in~$\mathcal{N}$ with rank below~$a$. Therefore, 
we can remove $N/56$ elements with rank below~$a$. In a similar manner, we can find
at least $N/56$ elements in~$\mathcal{N}$ with rank
above~$b$. Removal of these $2N/56 = N/28$ elements
in~$\mathcal{N}$ constitutes a type~A removal step.

Since the number of elements under consideration, i.e. $N$, 
decreases by a constant factor at each step, the algorithm
performs $O(\log n)$ type~A and type~B removal steps before we have
$N \le 64 \cdot \RUNS \cdot \log n$.  Since each block under 
consideration in a type~A removal step has size $\log n$, we can guarantee that each element in a
partitioning block
only needs to be selected as a partitioning element $O(1)$ times. This implies
that a total cost of $O(\log (\RUNS))$ \frag is incurred on each
element once we have that $N \le 64 \cdot \RUNS \cdot \log n$.

We now describe the final step of the algorithm.
At this point, the algorithm combines the last $\mathcal{N}$ elements
with all the short runs removed during its execution up to this point, forming the
set~$\mathcal{S}$. This set is the original elements
subjected to a series of type~A removals, each of which are balanced
and outside the rank interval $[a,b]$. Hence, the median
of~$\mathcal{S}$ is the global median. As
$|\mathcal{S}| = O(\RUNS \cdot \log n)$, we can find this median in
$O(\log(\RUNS \cdot \log n)) = O(\log (\RUNS) + \log\log n)$
\frag~\cite{afshani:fragile-ESA19}, which dominates the total
\frag of the algorithm.

We note that for $\RUNS = 2$, we can improve the above result to
$O(1)$ \frag as follows. Let the two runs be $R_1$ and $R_2$, with
$|R_1| \le |R_2|$.  Compare their middle elements $x$ and $y$ and
assume \ $x \le y$. Then the elements in the first half of $R_1$ are
below $n/2$ other elements, and hence are below the median. Similarly,
the elements in the last half of $R_2$ are above the median. Hence, we
can remove $|R_1|/2$ elements on each side of the median by removing
that many elements from one end of each run. The median of the
remaining elements is equal to the global median. By recursion, we in
$\log |R_1|$ steps end up with $R_1$ reduced to constant length. Then
$O(1)$ comparisons with the center area of $R_2$ will find the median.
Because both runs lose elements in each recursive step, both $x$ and
$y$ will be new elements each time. The total \frag of the
algorithm is therefore $O(1)$. \qed
\end{proof}

\begin{theorem}
\textsc{Minimum} has \frag $\Theta(\log (\INV))$.
\end{theorem}
\begin{proof}
Lower bound: For any $k$, consider the instances composed of
$\sqrt{k}$ elements in random order followed by $n - \sqrt{k}$
larger elements in sorted order. These instances have $\INV \le
k$. Finding the minimum is equal to finding the minimum on the first
$\sqrt{k}$ elements, which has a lower
bound~\cite{afshani:fragile-ESA19} of
$\log \sqrt{k} = \Omega(\log k)$ on its \frag.

For the upper bound, we will remove a subset~$I$ of size $O(\INV)$
which leaves a single sorted run~$R$. We can find the mininum in~$I$
in $O(\log(\INV))$ \frag by a tournament tree, which can then be
compared to the head of~$R$ for the final answer.

We find~$I$ and~$R$ in $O(1)$ \frag during a scan of the input as
follows, using~$R$ as a stack. For each new element~$e$ scanned, we
compare it to the current top element~$f$ of~$R$. If $e$ is larger, we
push~$e$ to the top of~$R$. If~$e$ is smaller, it forms an inversion
with~$f$, and we include $e$ in $I$. We also put a mark on~$f$. If an
element on the stack gets two marks, we pop it, include it in~$I$,
remove one of its marks (which will account for its inclusion in~$I$)
and move the second mark to the new top element~$f'$ of~$R$. If $f'$
now has two marks, this process continues until an element with only a
single mark is created (or the stack gets empty). An element is
compared with exactly one element residing earlier in the input (when
the element is scanned). To count comparisons with elements residing
later in the input, call such a comparison large or small, depending
on whether the other element is larger or smaller. It is easy to see
that elements on the stack always have between zero and one marks,
that an element with zero marks has participated in one large
comparison, and that an element with one mark has either participated
in at most two larger comparisons or one smaller comparison. Hence,
the \frag of the process is~$O(1)$. By the accounting scheme, $I$
is no larger than $\INV$ plus the number of marks, which is also at
most~$\INV$. \qed
\end{proof}

\begin{theorem}
\textsc{Median} has \frag $\Theta(\log (\INV))$.
\end{theorem}
\begin{proof}
As \textsc{Median} solves \textsc{Minimum} via padding with $n$
elements of value~$-\infty$, the lower bound follows from the lower
bound on~\textsc{Minimum}. For the upper bound, find~$R$ and $I$ as in
the upper bound for \textsc{Minimum}, sort~$I$ in
\frag~$O(\log(\INV))$ and use the algorithm for \textsc{Median}
for $\RUNS = 2$. \qed
\end{proof}

\begin{theorem}
\textsc{Sorting} has \frag $\Theta(\log (\INV))$.
\end{theorem}
\begin{proof}
The lower bound follows from the lower bound on~\textsc{Minimum}.
For the upper bound, find~$R$ and $I$ as in the upper bound for
\textsc{Minimum} and let each element recall its position in the
input. Divide the sorted sequence~$R$ into contiguous blocks of size
$|I|$ and let $R_i$ be the set of $i$'th elements of all blocks.
With the $i$'th element of~$I$ we perform an exponential search on
$R_i$, starting from the block where the element's position in the
input is. If the search moves a distance~$k$, the element from~$I$
participated in at least~$k$ inversions in the input, so
$k \le \INV$ and hence the incurred \frag for the element
is~$O(\log k) = O(\log(\INV))$. A \frag of~$O(1)$ is incurred on the
elements of~$R$, as each~$R_i$ is used once. After this, each
element from~$I$ knows its position in~$R$ within a window of
size~$|I|$. If the window of an element and a block overlaps, we
call the element and the block associated. Each block of~$R$ is
associated with at most~$|I| = O(\INV)$ elements, and each element
is associated with at most two blocks. For each block in turn, we
now sort its associated elements and merge them into the block
(except for tail elements overlapping the next block). The sorting
incurs~$O(\log(\INV))$ \frag, as does the merging if we use
exponential merging~\cite{afshani:fragile-ESA19}. We remove all
inserted elements from any association with the next block. Then we
continue with the next block. \qed
\end{proof}